\documentclass[preprint,floats,aps,epsfig,nofootinbib,amssymb]{revtex4-1}
\usepackage{mathrsfs}

\usepackage{slashed}
\usepackage{graphicx,color}
\usepackage{dcolumn}
\usepackage{bm}
\usepackage{subfig}
\usepackage{graphicx}
\usepackage{amssymb}
\usepackage{stackrel}
\usepackage{pgfplots}
\usetikzlibrary{positioning}
\usetikzlibrary{snakes}

\begin{document}

\title{Direct detections of Majorana dark matter in vector portal }

\author{Wei Chao$^1$}
\email{chaowei@bnu.edu.cn}
\affiliation{$^1$Center for Advanced Quantum Studies, Department of Physics, Beijing Normal University, Beijing, 100875, China}
\vspace{3cm}

\begin{abstract}

In this paper we investigate the direct detections of Majorana dark matter (MDM) in vector portal. Taking into account that the tree-level scattering cross sections in these models are either dark matter velocity suppressed or spin-dependent, we calculate radiative corrections to the spin-independent cross section in effective field theory approach. Wilson coefficients of effective MDM-quark interactions are calculated at the one-loop level, and the Wilson coefficient of the effective MDM-gluon interaction is derived at the two-loop level. Numerical results show that current constraints can rule out a narrow mass range of MDM when  tree-level contributions are considered, and the spin-independent cross section  from radiative corrections  is reachable by the current direct detection technique for  light MDM. 

\end{abstract}

\maketitle
\section{Introduction}
Various observations have confirmed the existence of dark matter in our universe, whose relic density, derived by measuring the cosmic microwave background, large scale structure and galaxy formation, is  about $0.1198\pm0.0033$~\cite{Aghanim:2018eyx}.  The standard model (SM) of particle physics contains no cold dark matter candidate, and the nature of dark matter remains elusive. There are many dark matter candidates with  masses ranging from $10^{-20}$ eV to $10^{55}$ GeV, of which the weakly interacting massive particle (WIMP)~\cite{Goldberg:1983nd,Ellis:1983ew,Jungman:1995df,Servant:2002aq,Cheng:2002ej,Bertone:2004pz} is well-motivated as it can naturally explain the observed relic density via the thermal freeze-out with its mass at the electroweak scale and with weak couplings  to the SM particles. 

There are three (direct or indirect) ways of detecting WIMPs in laboratories:  looking for the scattering between WIMPs and nucleon in underground laboratories by measuring the nuclear recoil energy in the kilo-electronvolt scale, detecting the flux of cosmic rays injected by the WIMP annihilations or decays with the help of satellites or telescopes, and  producing WIMPs at the Large Hardron Collider (LHC) where the signal of WIMP is missing transverse momentum or missing energy. Of these three efforts, the first detection method is most straightforward since the astrophysical sources of cosmic rays have not been clearly determined in indirect detection experiments,  and  LHC is actually a mediator machine in dark matter detections.  

Benefiting from technological advances, direct detection experiments such as LUX~\cite{Akerib:2013tjd}, PandaX-II~\cite{Tan:2016zwf}  and XENON1T~\cite{Aprile:2017iyp} have made tremendous strides in increasing precision and detecting  efficiency. In an ideal status, one can detecting arbitrarily small direct detection cross section by continuously increasing the exposure, however it is well-known that direct detection experiments will soon reach an irreducible background from coherent  elastic neutrino-nuclei scattering, the so-called ``neutrino floor"~\cite{Billard:2013qya}.  The current direct detection techniques will not be able to distinguish the signal of dark matter from that of neutrinos  if the signal lies below the neutrino floor. 
That is to say the neutrino floor is the border of new and ``old" direct detection techniques.   As a result, 
the precision calculation of the direct detection cross sections will be important, if one wants to examine as many dark matter models as possible with the help of current direct detection techniques.

In this paper, we study the direct detections (DD) of a vector portal Majorana dark matter (MDM)~\cite{Dudas:2009uq}. The vector mediator model is one of the simplest models, whose phenomenology has been widely studied in Refs.~\cite{An:2012va,Frandsen:2012rk,Dreiner:2013vla,Alves:2013tqa,Arcadi:2013qia,Lebedev:2014bba,Bell:2014tta,Alves:2015pea,DeSimone:2016fbz,Fairbairn:2016iuf,Cui:2017juz}. In this model $\chi$ is a Majorana dark matter, $V_\mu^{}$ is a vector mediator, whose mass may arise from the spontaneous breaking of certain $U(1)$ gauge symmetry, and  $V_\mu^{} $ may couple to the SM via a vector current or axial-vector current. In some  models the vector portal is in associated with the Higgs portal since the scalar that causes the spontaneous breaking of the new $U(1)$ gauge symmetry may mix with the SM Higgs. Here we assume the mixing is negligibly small, thus the scattering of $\chi$ off the nuclei is only mediated by the $V_\mu$. 
The direct detection cross section $\sigma$ is either spin-independent but suppressed by the dark matter velocity or spin-dependent~\cite{Fitzpatrick:2012ix,Anand:2013yka}.  As a result, the spin-independent cross section $\sigma_{SI}^{}$, generated at the loop level~\cite{Haisch:2013uaa,Crivellin:2014gpa,DEramo:2016gos,Crivellin:2014qxa,Bishara:2018vix,Li:2018qip,Sanderson:2018lmj,Hisano:2010fy,Hisano:2011cs,Ertas:2019dew,Ishiwata:2018sdi,Abe:2018emu,Chao:2018xwz}, turns out to be important as it may be still possible  to examine these models with the current DD technique if $\sigma_{SI}^{}$ lies above the neutrino floor.  We calculate effective operators for the evaluation of MDM-nucleon spin independent scattering cross section following~\cite{Hisano:2015bma,Hisano:2017jmz}. The  WIMP-gluon effective operator~\cite{Hisano:2010ct,Hisano:2015rsa} raising  at the two-loop level is also derived. Numerical simulations show that the $\sigma_{SI}^{}$ from radiative corrections is reachable by the current DD technique for light MDM. 

The remaining of the paper is organized as follows: In section II we give a brief introduction to the vector portal MDM model. Section III is focused on the calculation of Wilson coefficients of effective operators. Numerical results  are presented in section IV and the last part is concluding remarks.  Expressions of loop functions are listed in the appendix A. Nuclear form factors are given in the appendix B.
  
\section{The model}
In this section, we review the vector portal dark matter model. The simplified model contains  a Majorana fermion $\chi$ and a new vector boson $V_\mu$ in addition to the SM particles. The Lagrangian for $\chi$ can be written as
\begin{eqnarray}
{\cal L}_\chi^{} = {1\over 2 } \bar \chi i \slashed{\partial} \chi + {1\over 2} g_V^{} \bar \chi \gamma^\mu \gamma^5 V_\mu^{} \chi - {1\over 2} m_\chi^{} \bar \chi \chi
\end{eqnarray}
where $m_\chi^{}$ is the dark matter mass, $g_V^{} $ is the new gauge coupling. The ultraviolet completed model contains a U(1) gauge symmetry and a complex scalar $\Phi$($\equiv{1\over \sqrt{2}} (\phi + i G + v_\Phi^{})$) which is charged under the new U(1) and whose vacuum expectation value (VEV) $v_\Phi^{}$ leads to the spontaneous breaking of the new gauge symmetry and the origin of masses of $\chi$ and $V$. In this case there will be a new Yukawa interaction, ${1\over 2} y_\chi  \overline{\chi^C_A} \Phi \chi^{}_A + {\rm h.c.}$, where the subindex $A$ represents the chirality, and Eq.(1) needs to extended with the ${1\over 2} \bar \chi \phi \chi$ term.  In the case where $\phi$ mixies with the SM Higgs, there will be Higgs portal interactions.  Notice that the mixing is caused by the quartic term: $\lambda_{H\Phi} \Phi^\dagger \Phi H^\dagger H$. $\lambda_{H\Phi}$ should be small when  taking into account constraints from precision observables as well as Higgs measurements at the LHC. Here we assume the the mixing is negligible. 

For interactions of new gauge boson with the SM particles, $V_\mu$ may couple to vector bilinears or axial-vector bilinears, or both, depending on the U(1) charge settings of the SM fermions. For example, new gauge interactions are $g_V^{} \bar f \gamma^\mu f $ in $U(1)_{B-L}$~\cite{Mohapatra:1980qe}, $U(1)_{B+L}$~\cite{Chao:2016avy} and $U(1)_{B_i-L_j}^{}$ models.  While in the model where two chiral components carry opposite U(1) charges, new gauge interactions will be $g_V^{} \bar f \gamma^\mu \gamma^5 f $.  If only one certain chirality component carries non-zero charge, new gauge interactions will be $g_V \bar f \gamma^\mu_{} P_A V_\mu^{} f  $~\cite{Chao:2017rwv}, where $P_A=P_L^{}$ or $P_R^{}$. In this paper we carry out model  independent study and investigate the DD cross section of vector current and axial-vector current scenarios separately. Interactions of $V_\mu$ with quarks are then
\begin{eqnarray}
&&{\cal L}_I^q \in \zeta g_V^{}  \bar q \gamma^\mu q V_\mu^{} \; , {\rm \text(scenario~A)} \\
&&{\cal L}_{II}^q \in \zeta g_V^{}  \bar q \gamma^\mu \gamma^5 q V_\mu^{} \; ,  {\rm \text(scenario~B)}
\end{eqnarray}
where $\zeta$ is  the U(1) hyper-charge of the quark $q$. Free parameters in these models are thus $m_\chi, ~m_V, ~g_X $and $ \zeta$.

\section{Effective operators}
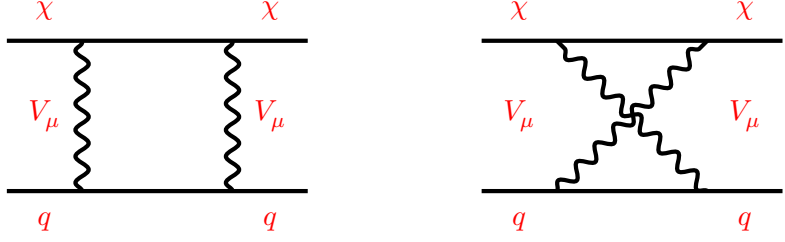
\begin{figure}
\begin{center}
\begin{tikzpicture}
\draw[-,ultra thick] (-1,0)--(3,0);
\draw [-,snake=snake, ultra thick] (0,0) -- (0,2);
\draw [-,snake=snake, ultra thick] (2,0) -- (2,2);
\draw[-,ultra thick] (-1,2)--(3,2);
\node[red, thick] at (-0.5,-0.4) {$q$};
\node[red, thick] at (2.5,-0.4) {$q$};
\node[red, thick] at (-0.5,1.0) {$V_\mu^{}$};
\node[red, thick] at (2.5,1.0) {$V_\mu^{}$};
\node[red, thick] at (-0.5,2.4) {$\chi$};
\node[red, thick] at (2.5,2.4) {$\chi$};
\end{tikzpicture}
\hspace{2cm}
\begin{tikzpicture}
\draw[-,ultra thick] (-1,0)--(3,0);
  \draw [-,snake=snake, ultra thick] (0,0) -- (2,2);
   \draw [-,snake=snake, ultra thick] (2,0) -- (0,2);
   \draw[-,ultra thick] (-1,2)--(3,2);
   \node[red, thick] at (-0.5,-0.4) {$q$};
\node[red, thick] at (2.5,-0.4) {$q$};
\node[red, thick] at (-0.5,1.0) {$V_\mu^{}$};
\node[red, thick] at (2.5,1.0) {$V_\mu^{}$};
\node[red, thick] at (-0.5,2.4) {$\chi$};
\node[red, thick] at (2.5,2.4) {$\chi$};
\end{tikzpicture}
    \caption{Box diagrams for the effective quark-WIMP interactions}\label{boxd}
\end{center}
\end{figure}

In this section we calculate effective operators relevant for the spin-independent scattering cross section of $\chi$ with a nucleon. Following Refs.~\cite{Hisano:2015bma,Hisano:2017jmz} , we write down the effective $\chi$-quark interactions in terms of the higher-dimensional operators:
\begin{eqnarray}
{\cal L}_{\rm eff}^{} = \kappa_0^{}  \bar \chi \gamma^\mu \gamma^5 \chi \bar q \Gamma q + \sum_{p=q,g}  {1\over 2}\kappa_{1p} \bar \chi \chi {\cal O}_s^p + {1\over 2}\kappa_{2q}^{}  \bar \chi i \partial^\mu \gamma^\nu \chi {\cal O}_{\mu \nu}^q + {1\over 2}\kappa_{3q}\bar \chi i \partial^\mu i \partial^\nu \chi {\cal O}_{\mu \nu}^q \ \; , 
\end{eqnarray}
where $\Gamma=\gamma^\mu$ for scenario A,  $\Gamma=\gamma^\mu \gamma^5$ for scenario B, $\kappa_{ip}^{}$ are wilson coefficients, ${\cal O}_s^q=m \bar q q $, ${\cal O}_s^g=-{9\over 8}\alpha_s^{} G^{A \mu \nu}_{} G_{\mu \nu}^A$ and ${\cal O}_{\mu \nu}^q$ are twist-2 operators, defined by
\begin{eqnarray}
{\cal O}_{\mu\nu}^q = {1\over 2 } \bar q \left( \partial_\mu \gamma_\nu + \partial_\nu \gamma_\mu - {1\over 2} g_{\mu\nu} \slashed{\partial} \right) q \; .
\end{eqnarray}
These effective operators are defined at the mass scale of $V_\mu$, which is assumed to be heavier than all the SM particles. In the following, we calculate  Wilson coefficients at the leading order. 

The leading contribution to the Wilson coefficient $\kappa_0$ arises from the tree level diagram by exchanging the vector boson $V_\mu^{}$, 
\begin{eqnarray}
\kappa_0^{} = {\zeta g_V^2 \over 2 m_V^2} \; ,
\end{eqnarray}
where $m_V^{}$ is the mass of $V_\mu^{} $.

$\kappa_{1,2,3}^q$ arise from the box diagrams shown in the Fig.~\ref{boxd}. We calculate the diagrams in the zero-momentum transfer limit, and expand the amplitude in term of the quark momentum, which is non-relativistic, then we  decompose results into effective operators following~\cite{Abe:2018emu}. For the scenario A, the relevant Wilson coefficients are 
\begin{eqnarray}
\kappa_{A1q}^{}
&=&- {m_\chi^{} \over 2}\alpha_V^2    \left\{ 3X_2(m_\chi^2, m_V^2, 0, m_\chi^2)+Y_2(m_\chi^2, m_V^2, 0, m_\chi^2) +12 Z_{001}^{}(m_\chi^2, m_V^2, m_\chi^2)\right.\nonumber \\&&\left. +6 Z_{00}^{}(m_\chi^2, m_V^2, m_\chi^2) + m_\chi^2\left[3  Z_{11}^{}(m_\chi^2, m_V^2, m_\chi^2) + 2 Z_{111}^{}(m_\chi^2, m_V^2, m_\chi^2) \right] \right\} \\
\kappa_{A2q}^{} 
&=& -2\alpha_V^2 \left\{2Z_{00}^{}(m_\chi^2, m_V^2, m_\chi^2) +8Z_{001}(m_\chi^2, m_V^2, m_\chi^2) + m_\chi^2 \left[Z_{11}^{}(m_\chi^2, m_V^2, m_\chi^2)\right. \right.\nonumber \\ && \left.\left.+ Z_{111}^{}(m_\chi^2, m_V^2, m_\chi^2) \right]-X_2^{}(m_\chi^2, m_V^2, 0, m_\chi^2) -  Y_2^{} (m_\chi^2, m_V^2, 0, m_\chi^2) \right\} \\
\kappa_{A3q}^{} &=& -2\alpha_V^2 \left[ 2Z_{11}^{}(m_\chi^2, m_V^2, m_\chi^2) + Z_{111}^{}(m_\chi^2, m_V^2, m_\chi^2) \right]
\end{eqnarray}
where $\alpha_V=g_V^2 /4\pi$. The definitions of the loop functions and their explicit expressions  are given in the appendix A. Loop functions are evaluated with the help of the Package-X~\cite{Patel:2015tea,Patel:2016fam}.

For scenario B the Wilson coefficients are 
\begin{eqnarray}
\kappa_{B1q}^{} &=&+{m_\chi \over 2 } \alpha_V^2 \left\{  9 X_2^{}(m_\chi^2, m_V^2, 0, m_\chi^2)+3 Y_2^{}(m_\chi^2, m_V^2, 0, m_\chi^2)-6Z_{00}^{}(m_\chi^2, m_V^2, m_\chi^2 )\right.\nonumber \\
&&\left.+13 Z_{001}^{}(m_\chi^2, m_V^2,  m_\chi^2) -m_\chi^2 \left[3Z_{11}^{}(m_\chi^2, m_V^2, m_\chi^2) - 2Z_{111}^{}(m_\chi^2, m_V^2, m_\chi^2)\right]\right\} \; ,  \\
\kappa_{B2q}^{} &=& -2 \alpha_V^2 \left\{ 2Z_{00}^{}(m_\chi^2, m_V^2, m_\chi^2) -8Z_{001}^{}(m_\chi^2, m_V^2, m_\chi^2) + m_\chi^2 \left[Z_{11}^{}(m_\chi^2, m_V^2, m_\chi^2)\right.\right. \nonumber \\ &&\left.\left.- Z_{111}^{}(m_\chi^2, m_V^2, m_\chi^2) \right]-X_2^{} (m_\chi^2, m_V^2, 0, m_\chi^2)- Y_2^{}(m_\chi^2, m_V^2, 0, m_\chi^2)\right\} \; ,  \\
\kappa_{B3q}^{} &=& -2 \alpha_V^2 \left[2 Z_{11}^{}(m_\chi^2, m_V^2, m_\chi^2) - Z_{111}^{}(m_\chi^2, m_V^2, m_\chi^2) \right]\; .
\end{eqnarray}

\begin{figure}[t]
\begin{center}
\begin{tikzpicture}
\draw[-,ultra thick] (-1,2)--(3,2);
\draw [-,snake=snake, ultra thick] (0.5,0.3) -- (0.5,2);
\draw [-,snake=snake, ultra thick] (1.5,0.3) -- (1.5,2);
\draw [snake=coil, segment length=4pt, ultra thick] (-1,0) -- (0,0);
\draw [snake=coil, segment length=4pt, ultra thick] (2,0) -- (3,0);
\draw (1,0) ellipse (28pt and 10 pt);
\node[red, thick] at (-0.5,-0.4) {$G$};
\node[red, thick] at (2.5,-0.4) {$G$};
\node[red, thick] at (0,1.0) {$V_\mu^{}$};
\node[red, thick] at (2,1.0) {$V_\mu^{}$};
\node[red, thick] at (-0.5,2.4) {$\chi$};
\node[red, thick] at (2.5,2.4) {$\chi$};
\end{tikzpicture}
\hspace{2cm}
\begin{tikzpicture}
\draw[-,ultra thick] (-1,2)--(3,2);
\draw [-,snake=snake, ultra thick] (0.5,0.3) -- (0.5,2);
\draw [-,snake=snake, ultra thick] (1.5,-0.3) -- (1.5,2);
\draw [snake=coil, segment length=4pt, ultra thick] (-1,0) -- (0,0);
\draw [snake=coil, segment length=4pt, ultra thick] (2,0) -- (3,0);
\draw (1,0) ellipse (28pt and 10 pt);
\node[red, thick] at (-0.5,-0.4) {$G$};
\node[red, thick] at (2.5,-0.4) {$G$};
\node[red, thick] at (0,1.0) {$V_\mu^{}$};
\node[red, thick] at (2,1.0) {$V_\mu^{}$};
\node[red, thick] at (-0.5,2.4) {$\chi$};
\node[red, thick] at (2.5,2.4) {$\chi$};
\end{tikzpicture}
    \caption{Two-loop Feynman diagrams for the effective gluon-WIMP interactions.}\label{twoloop}
\end{center}
\end{figure}
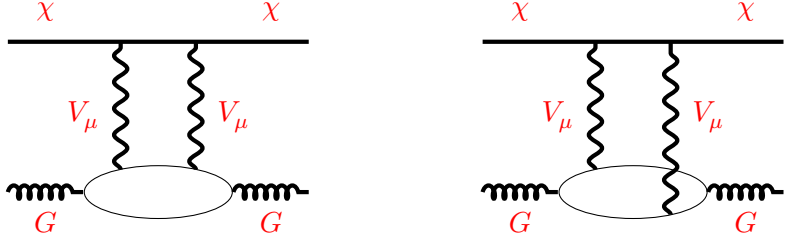

The effective MDM-gluon interactions arise at the two-loop level. Relevant Feynman diagrams are given in the Fig.~\ref{twoloop}. In this paper we only take into account the effect of twist-0 operator while neglecting that of higher twist operators.  One loop correction to the two-point function of gauge boson in the gluon background field has been calculated in Ref.~\cite{Abe:2015rja} by taking the Fock-Schwinger gauge~\cite{Novikov:1983gd,Shtabovenko:2016sxi} for the gluon field,  which,  mapped into our cases, can be written as 
\begin{eqnarray}
i\Pi_{VV}^{(f) \alpha \beta} =-{1\over 3} {i \zeta^2 g_s^2 \over 16 \pi } G^a_{\mu \nu} G^{a \mu \nu} \left({  g_V^2 \over q^2 } g^{\alpha \beta} -  {g_V^2 \over q^4 } q^\alpha q^\beta \right) \label{ggvv}
\end{eqnarray} 
where $g_s$ is the coupling of the strong interaction, $f$ indicates the flavor running in the fermion loop, $q$ is the momentum of gauge boson.  Notice that Eq.~(\ref{ggvv}) is universal for both the scenario A and scenario B.

With the help of Eq. (\ref{ggvv}), one can write down the Wilson coefficient of the effective $\chi$-gluon  operator as 
\begin{eqnarray}
\kappa_{Ag(Bg)}^{}&=& {\alpha_V^2 \over 54} \zeta^2 n_f^{}  m_\chi^{}  \left( 6 X_2^{}(m_\chi^2, m_V^2, 0, m_\chi^2) + 2 Y_2^{}(m_\chi^2, m_V^2, 0, m_\chi^2)\right. \nonumber \\ &&\left. + 6 Z_{001}^{}(m_\chi^2, m_V^2, m_\chi^2) -6Z_{00}^{}(m_\chi^2, m_V^2, m_\chi^2) + m_\chi^2 Z_{11}^{}(m_\chi^2, m_V^2, m_\chi^2) \right)
\end{eqnarray}
where $n_f $ is the number of quarks that carry nonzero U(1) charge. 

Wilson coefficients given above are matched to the simplified model at $\mu\approx m_V$. The energy scale for the DM direct detections is about the nuclear energy scale.  It has been shown in Ref.~\cite{DEramo:2016gos} that effects  from the running of  renormalization group equations  might be sizable in certain vector-portal scenario.   We use the public code RUNDM~\cite{DEramo:2016gos} to evolve the running of $\kappa_0$ and evaluate running effects of other Wilson coefficients  following Refs.~\cite{Hisano:2015rsa,DEramo:2014nmf,Hill:2014yka,Hill:2014yxa,Mohan:2019zrk}.

\section{Results}

In this section we present results for the MDM-nucleon scattering cross section. We start by determining the $g_V$ using the observed relic density. As was mentioned in the previous section, there are four parameters in the vector portal, which are all relevant for both the relic abundance and the direct detection cross section.  The thermal relic abundance is determined by the following processes: $\bar \chi \chi \to \bar f f $ and $\bar \chi \chi \to V_\mu V^\mu$, where $f$ indicates the SM fermion. The first channel depends on the parameter $\zeta$, while the second channel does not.  For $m_\chi < m_V$, the channel $\bar \chi \chi \to V_\mu V^\mu$ is kinematically forbidden, and thus the combination $ {\zeta} g_V^{2} $ can be determined in this mass range.

\begin{figure}
\includegraphics[width=0.45\textwidth]{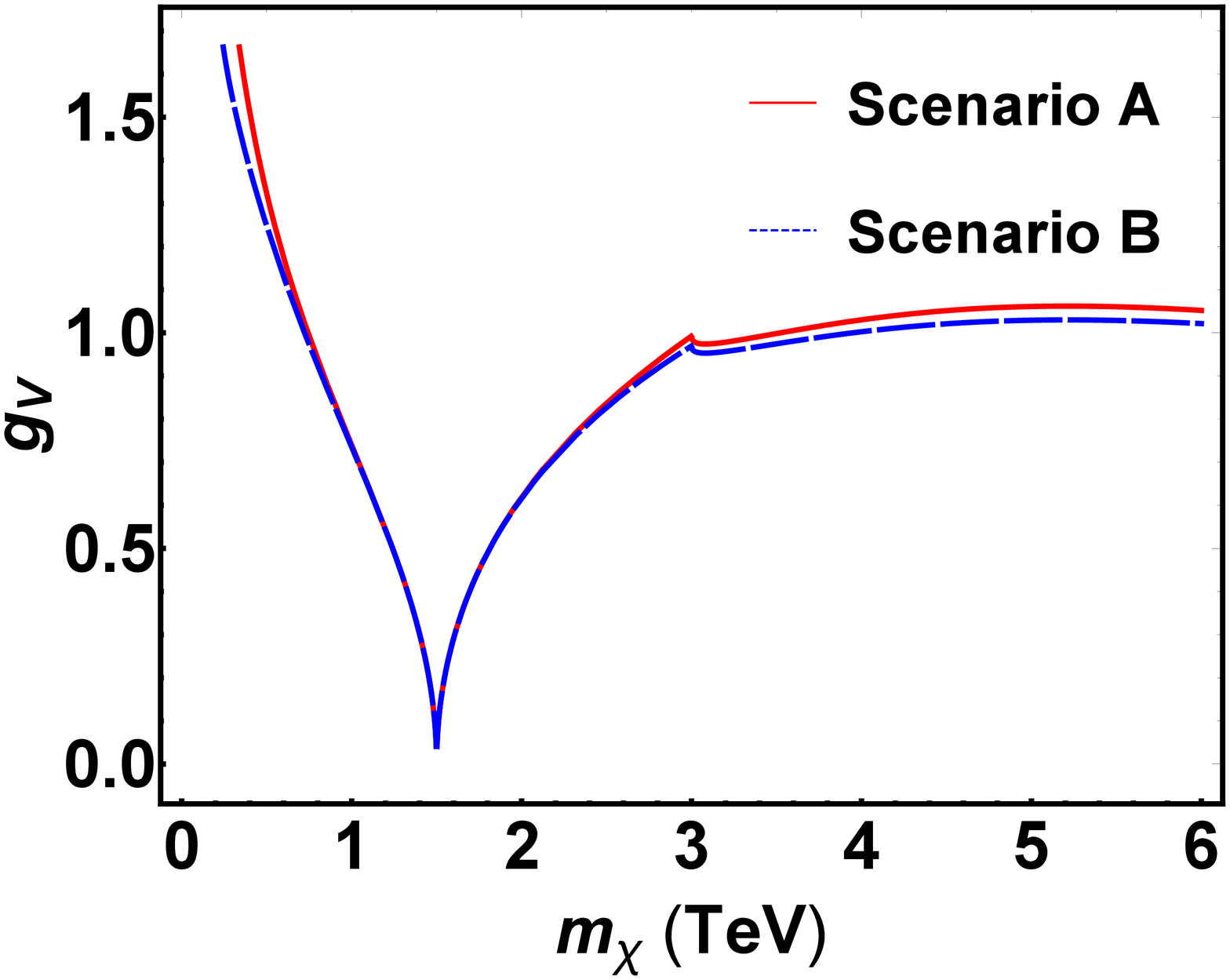}
\hspace{0.5cm}
\includegraphics[width=0.45\textwidth]{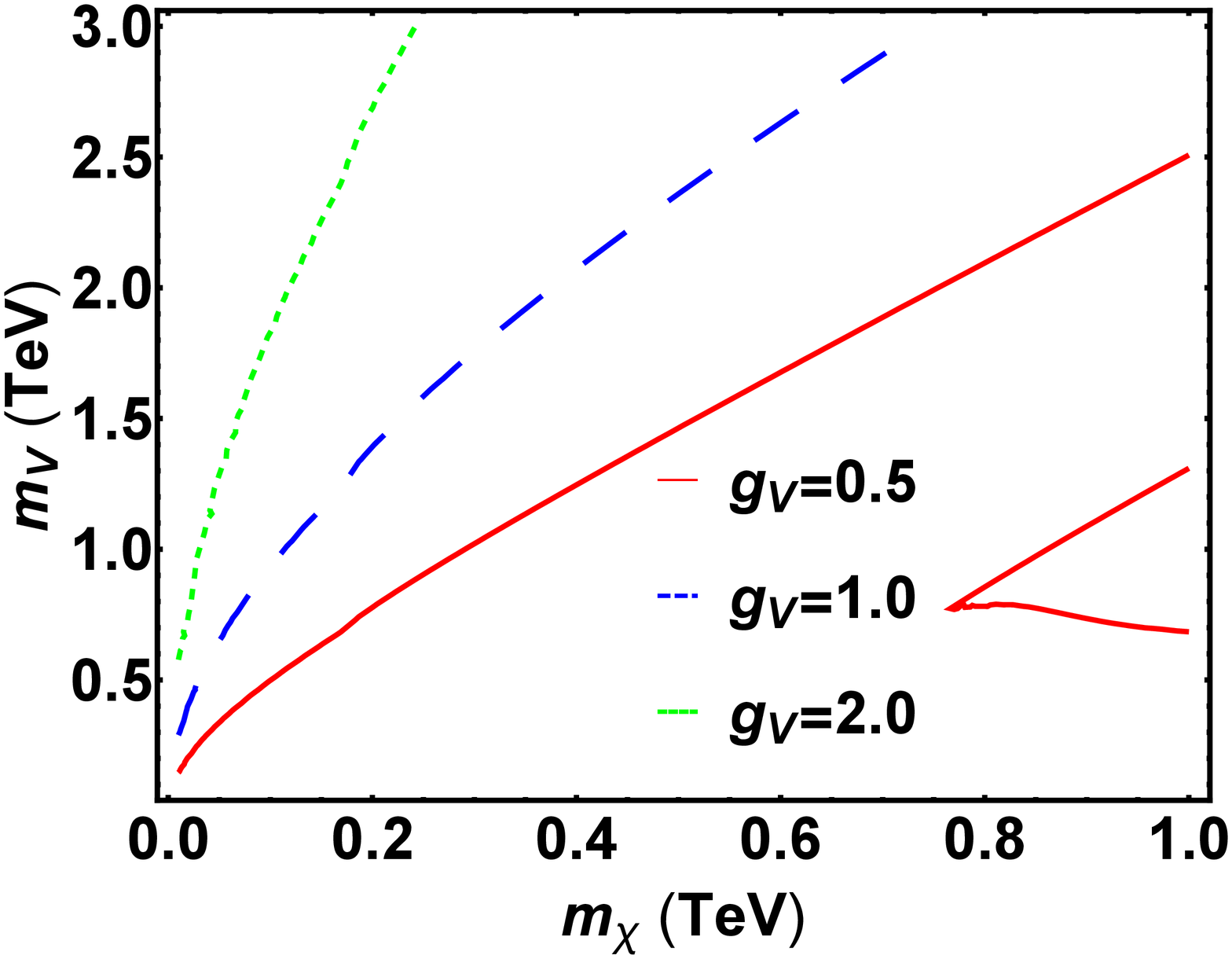}
\caption{ Left panel: New gauge coupling as the function of the dark matter mass by setting $m_V=3~{\rm TeV}$ and $\zeta=1$, constrained by the observed relic density. The solid and dashed lines correspond to the scenario I and scenario II respectively. Right panel: Contours of $g_V$ in the $m_\chi-m_V$ plane by setting $\zeta=1$.
} \label{fig:phase}
\end{figure}

We show in the Fig.~\ref{fig:phase}, the new gauge coupling $g_V$ as the function of the dark matter mass $m_\chi$ determined by the observed relic abundance $\Omega h^2 =0.1198$,  by setting $m_V=3~{\rm TeV}$ and $\zeta=1$. The solid line and dashed line correspond to cases of scenario A and scenario B, respectively.  The first dip of the plot appears at $m_\chi \sim m_V/2$, where the annihilation $\bar \chi \chi \to \bar f f$ is resonantly enhanced. The second dip of the plot at $m_\chi\sim m_V$ is due to the opening of the channel $\bar \chi \chi \to V_\mu^{} V_{}^\mu$.  Notice that  $g_V$ becomes almost scenario independent as $m_\chi \sim m_V$. We show in the right panel of the Fig.~\ref{fig:phase} contours of $g_V$ in the $m_\chi-m_V$ plane by setting $\zeta=1$. Notice that ${\cal O}(g_V) \sim 1$ for $m_V\sim m_\chi \sim {\cal O}(1) $ TeV. 

\begin{figure}
\centering
\includegraphics[width=0.45\textwidth]{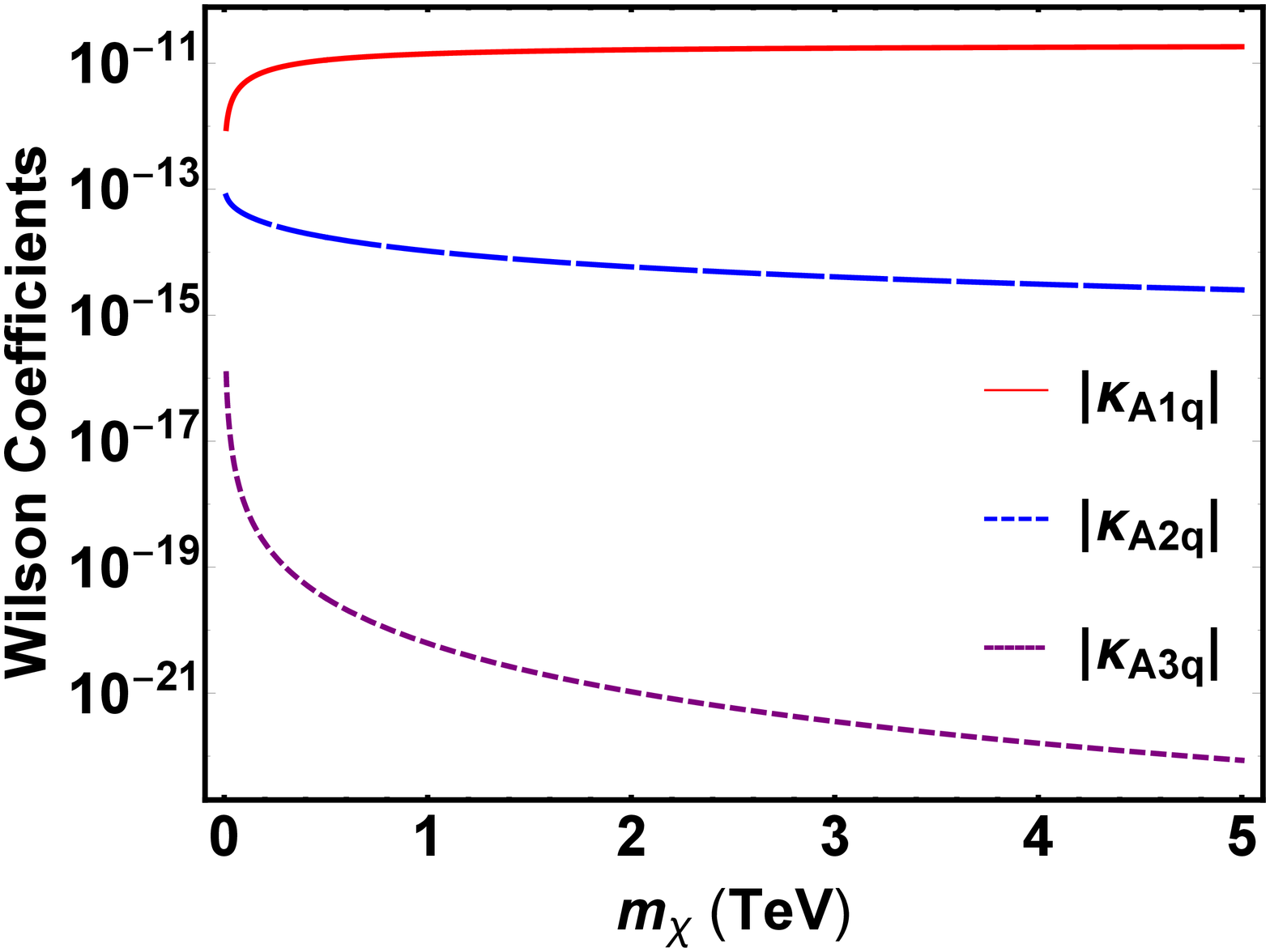}
\hspace{0.5cm}
\includegraphics[width=0.45\textwidth]{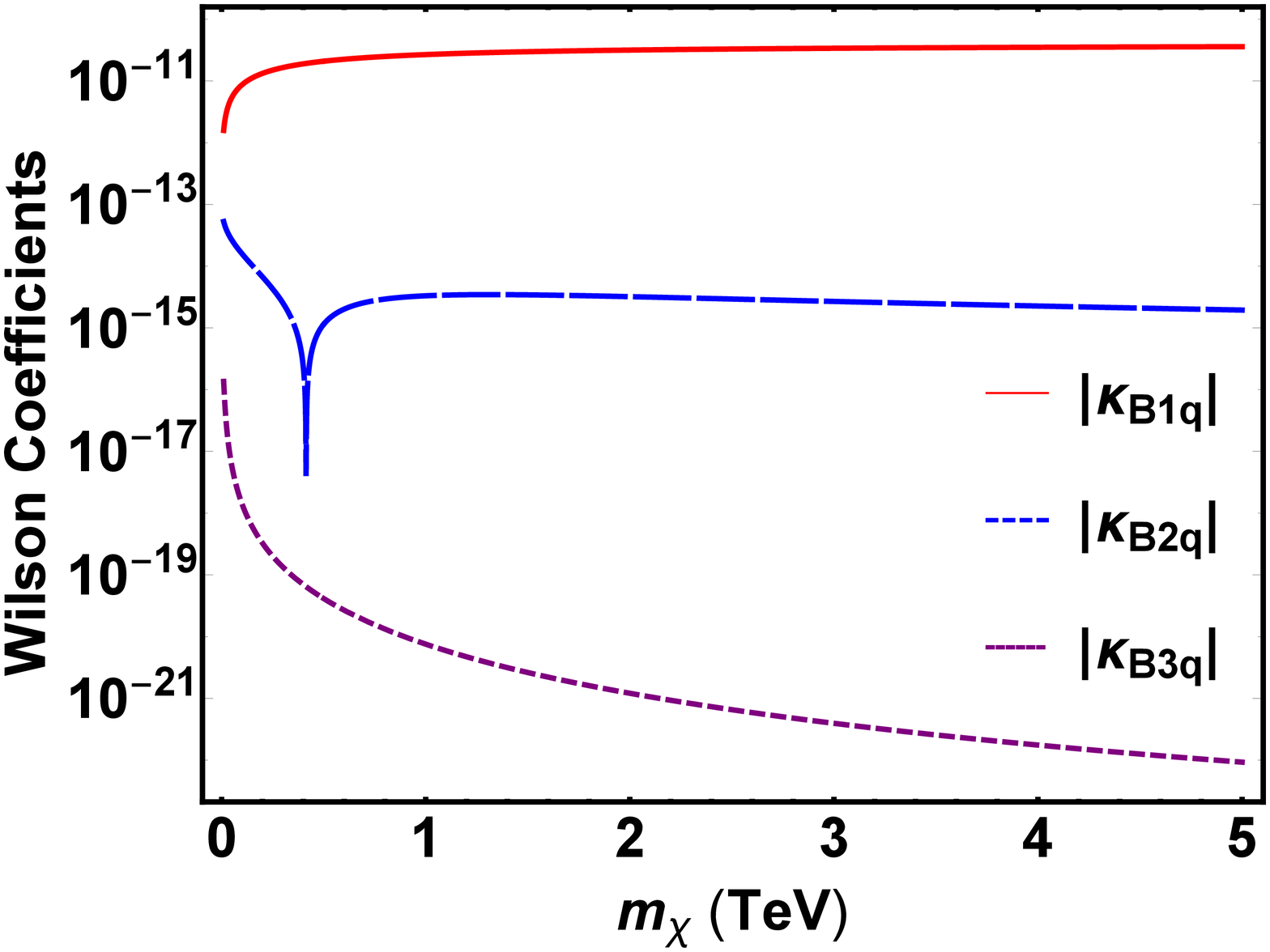}
\caption{ Wilson coefficients as the function of the dark matter mass by setting $\zeta=g_V=1$ and $m_V=1~{\rm TeV}$ for scenario A (left-panel) and scenario B (right-panel).
} \label{wilsonco}
\end{figure}

We show in the Fig.~\ref{wilsonco}  Wilson coefficients as the function of the dark matter mass $m_\chi$ by setting $g_V^{}=\zeta=1$ and $m_V=1~{\rm TeV}$.  The plot in the left-panel and right-panel correspond to cases of scenario A and B respectively.  As can be seen, the Wilson coefficient of the scalar type interaction is largest and is comparable to that in pseudo-scalar portal model~\cite{Abe:2018emu}.   Wilson coefficients in scenario B is similar to these in scenario A except  $\kappa_{B2q}^{}$ at $m_\chi\sim m_V$, which is due to the cancellation of various contributions.  The Wilson coefficient $\kappa_g $ is of the order ${\cal O} (10^{-11})$ by setting $g_V=\zeta=1$.

\begin{figure}
\includegraphics[width=0.6\textwidth]{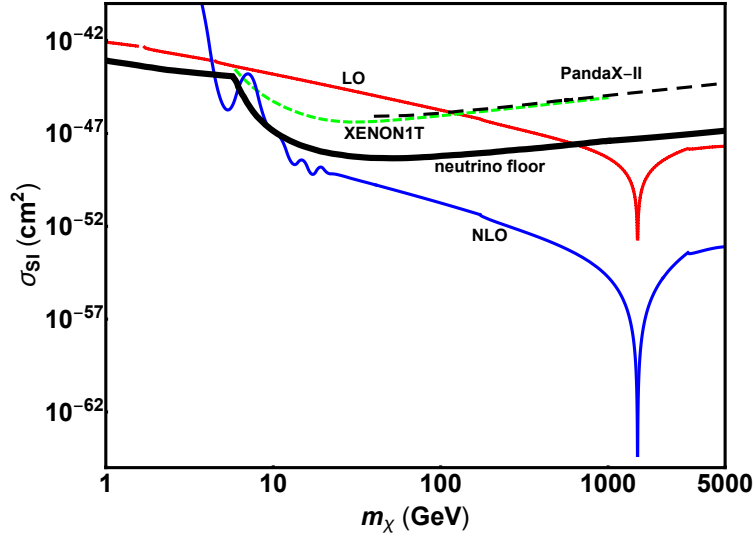}
\caption{Direct detection cross section as the function of the dark matter mass by setting $\zeta=1$ and $m_V=3~{\rm TeV}$ for scenario A. The red solid and blue dashed lines correspond to the LO and NLO contributions respectively. The black dashed and green dashed lines are separately constraints of Pandax-II and XENON1T. } \label{fig:finalA}
\end{figure}

Now we calculate the MDM-nucleon scattering cross section. For scenario A, there is velocity suppressed spin-independent scattering cross section at the leading order, 
\begin{eqnarray}
\sigma_{\rm SI}^{\rm LO} \approx {36\pi\zeta^2 \alpha_V^2  \mu^2  v^2  \over  m_V^4  }   \; ,
\end{eqnarray}
where $\mu$ is the reduced mass of $\chi$ and nucleon system, $v$ is the velocity of the dark matter. Effective interactions given in the Eq. (4) contribute to the scattering cross section at the next-to-leading order, 
\begin{eqnarray}
\sigma_{\rm SI}^{\rm NLO} ={\mu^2 m_N^2\over \pi} \left[ \sum_{q} \kappa_{A1q} f_{Tq}^N + \kappa_g f_{T_g}^N + {3\over 4} \sum_{q} (m_\chi^{} \kappa_{A2q}^{}  + m_\chi^2 \kappa_{A3q}^{} ) (q^N(2) +\bar q^N (2)) \right]^2  \label{xxx}
\end{eqnarray}
where  $N$ stands for $(p,~n)$ with $m_N$ its mass, $f_{T_q}^N$ is the quark matrix element defined by $\langle N|m_q \bar q q|N\rangle =m_N f_{T_q}^N$, $f_{Tg}^N$ is the gluon matrix element defined by $-{9\alpha_s\over 8\pi}\langle N|G^a_{\mu\nu} G^{a\mu\nu }|N\rangle =m_N f_{T_g}^N$, $q^N(2)$ and $\bar q^N(2)$ are second moments for quark distribution functions of $N$. Numerical values of $f_{T_q}^N$, $f_{T_g}^N$ as well as $q^N(2)$ and $\bar q^N(2)$  are listed in the appendix B. 
Notice that there is no interference between the leading oder and the next to leading oder contributions. As a result, the total cross section for scenario A can be written as $\sigma_{\rm SI}^{\rm tot} =\sigma_{\rm SI}^{\rm LO} + \sigma_{\rm SI}^{\rm NLO}$. 

\begin{figure}
\includegraphics[width=0.45\textwidth]{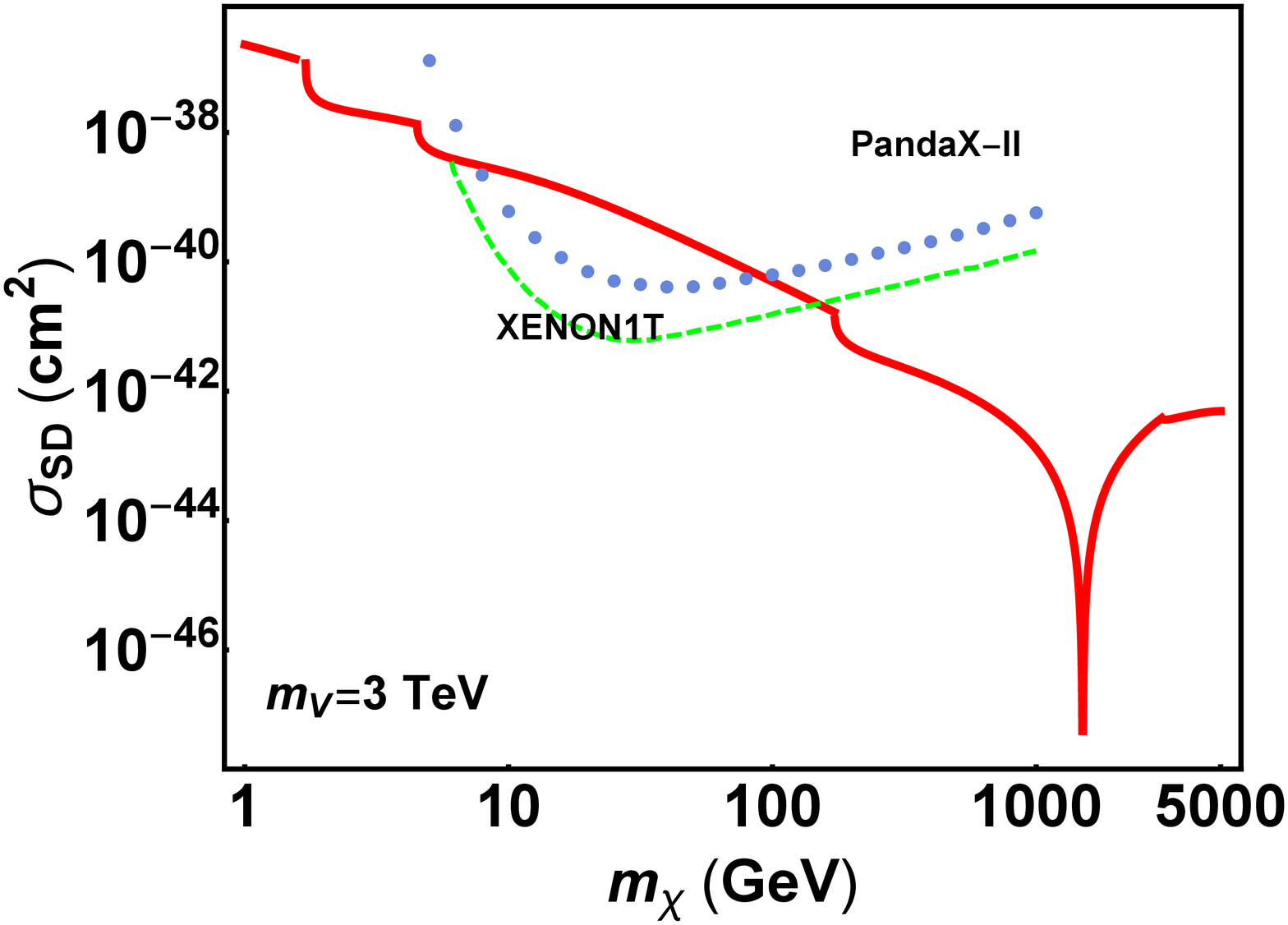}
\hspace{0.5cm}
\includegraphics[width=0.45\textwidth]{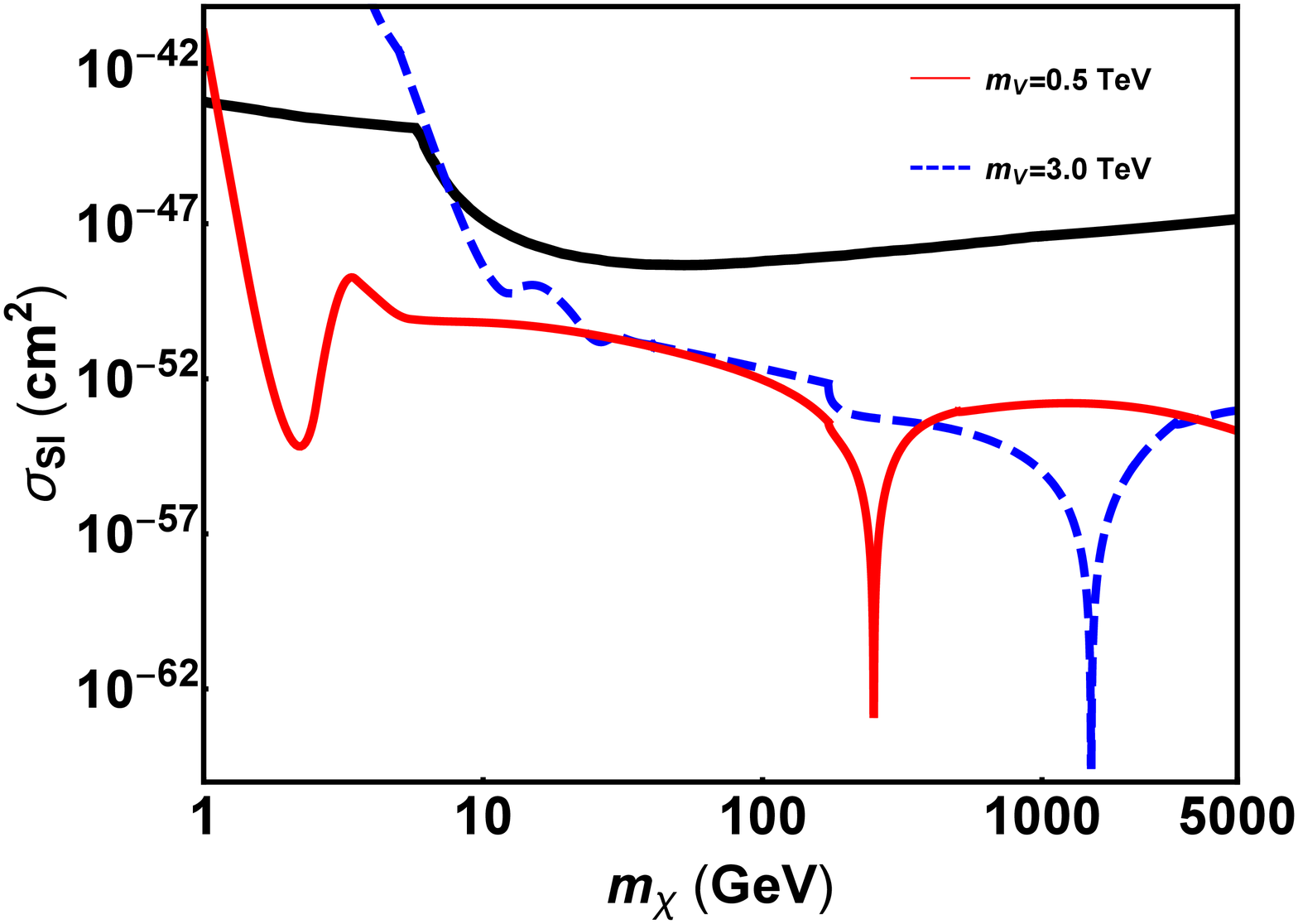}
\caption{Left-panel: Spin-dependent $\chi-$neutron scattering  cross section as the function of the dark matter mass by setting $\zeta=1$ and $m_V=3~{\rm TeV}$ for scenario B. The red  and blue dotted lines are constraints of XENON1T and PandaX-II, respectively. Right-panel: spin-independent $\chi-$neutron scattering cross section as the function of $m_\chi^{} $ by setting $\zeta=1$ and $m_V=3~{\rm TeV}$ for scenario B.  } \label{fig:finalB}
\end{figure}

For scenario B, the $\chi$-nucleon scattering cross section is spin-dependent at the leading order, 
\begin{eqnarray}
\sigma_{\rm SD}^{\rm LO} ={64 \pi}\zeta^2 \alpha_V^2 {\mu^2 \over m_V^4}\left(\sum_q \Delta_q^{N}\right)^2 J_N^{} (J_N^{} +1)   
\end{eqnarray}
where  $\Delta_q^N$ is the spin fraction of quark $q$, defined by $2\Delta_q^N s^\mu = \langle N |\bar q \gamma^\mu_{} \gamma^5_{} |N\rangle$ with $s^\mu $ the nucleon spin four-vector, $J_N$ is the angular momentum of the nucleon.  $\Delta_q^N$ are measured in DIS and one has $\Delta_u^p=0.77$, $\Delta_d^p=-0.47$ and $\Delta_s^p=-0.15$~\cite{Adams:1995ufa}. The $\chi$-nucleon scattering cross section at the next-to-leading order can be spin independent, whose expression is the same as eq. (\ref{xxx}), up to replacements, $\kappa_{Aiq} \to \kappa_{Biq}$ $(i=1,2,3)$.

As an illustration, we show in the Fig.~\ref{fig:finalA} the direct detection cross section as the function of the dark matter mass $m_\chi$ for scenario A, by setting $m_V=3~{\rm TeV}$ and $\zeta=1$. The red solid and blue dashed lines correspond to  $\sigma_{\rm SI}^{\rm LO}$ and $\sigma_{\rm SI}^{\rm NLO}$,  respectively. The black dashed and green dashed lines are separately constraints  given by the PandaX-II~\cite{Tan:2016zwf} and XENON1T~\cite{Akerib:2013tjd} experiments. One can conclude from the plot that even though the  $\sigma_{\rm SI}^{\rm LO}$ is suppressed by the dark matter velocity it is still sizable and the current constraints have excluded the low dark matter mass region ($6~{\rm  GeV}  m_\chi<120~{\rm GeV}$). The next-to-leading contribution is reachable by the current DD technique for light MDM, and it is about $5$ orders smaller than the LO term for heavy MDM.  The reason that $\sigma^{\rm NLO}$ being sensitive to light MDM is that $\sigma^{\rm NLO}$ is proportional to $g_V^8$ while $\sigma^{\rm LO}$ is proportional $g_V^4 $. As a result, a slight increase of $g_V$ may enhance the $\sigma^{\rm NLO}$ obviously. 

We show in the left-panel of the Fig.~\ref{fig:finalB} the spin-dependent $\chi$-neutron scattering cross section as the function of $m_\chi^{}$ for scenario B, by setting $\zeta=1$ and $m_V=3~{\rm TeV}$. The blue and green dotted lines are constraints of PandaX-II~~\cite{Fu:2016ega}  and XENON1T~\cite{Aprile:2019dbj}   experiments, respectively. One can see that the mass region $(8,~90)~{\rm GeV}$ is already excluded.   We show in the right-panel of the Fig.~\ref{fig:finalB} the spin-indpendent cross section as the function of $m_\chi$ for scenario B. The solid and dashed lines correspond to $m_V=3~{\rm TeV}$ and $0.5$ TeV, respectively. The black solid line is the neutrino floor. It is clear that the  spin-independent cross section is too small to be detected in the near future for heavy MDM, but it is detectable for light MDM.  $\sigma^{\rm NLO}$ is a good supplement to the $\sigma^{\rm SD} $ in detecting light MDM. 

\section{Conclusion}

In this paper we have discussed the direct detection cross section of Majorana dark matter $\chi$ in the vector portal.  At the leading order, the  cross section is either velocity suppressed or spin-dependent, and current constraints given by XENON1T and PandaX-II experiments can only rule out a narrow mass range depending on the inputs. Future direct detection experiments may improve the detection sensitivity to higher level. Next-to-leading order corrections may turn out to be important.  We have derived the effective $\chi$-quark interactions at the one-loop level and the effective $\chi$-gluon interaction at the two-loop level. Our numerical results show that the next-to-leading order corrections to the SI cross section is reachable by the current detecting technique for light MDM.  Notice that we did not consider constraints of LHC~\cite{Aaboud:2019yqu} since we are focused on DDs in underground laboratories in this paper, however constraints from collider are definitely important when considering a specific vector portal MDM model.

\begin{acknowledgments}
This work was supported by the National Natural Science Foundation of China under grant No. 11775025 and the Fundamental Research Funds for the Central Universities under grant No. 2017NT17.
\end{acknowledgments}

\appendix

\section{Integrations}

We list in this appendix definitions of integration used in this paper,  given by~\cite{Abe:2015rja}:
\begin{eqnarray}
 \int{d^4k\over(2\pi)^4}{1\over [(p+k)^2-M_\chi^2]k^2 [k^2-m_V^2]}&=&{i\over 16\pi^2} X_2(p^2,M_\chi^2, 0, m_V^2) \\
\int{d^4k\over(2\pi)^4}{k_\mu\over [(p+k)^2-M_\chi^2]k^2 [k^2-m_V^2]}&=&{i\over 16\pi^2} p_\mu Y_2(p^2,M_\chi^2, 0, m_V^2)\\
 \int{d^4k\over(2\pi)^4}{k_\mu k_\nu\over [(p+k)^2-M_\chi^2]k^4 [k^2-m_V^2]}&=&{i\over 16\pi^2}\left[ p_\mu p_\nu Z_{11}(p^2,M_\chi^2, m_V^2) +g_{\mu\nu} Z_{00} (p^2,M_\chi^2,m_V^2)\right ] \\ 
 \int{d^4k\over(2\pi)^4}{k_\mu k_\nu k_\sigma\over [(p+k)^2-M_\chi^2]k^4 [k^2-m_V^2]}&=&{i\over 16\pi^2}\left[ p_\mu p_\nu p_\sigma Z_{111}(p^2,M_\chi^2, m_V^2)  \right. \nonumber \\ &&\left. +(g_{\mu\nu}p_\sigma + g_{\mu \sigma } p_\nu + g_{\nu \sigma} p_\mu) Z_{000} (p^2,M_\chi^2,m_V^2)\right]
\end{eqnarray}
These integrations are evaluated  using package-X~\cite{Patel:2015tea,Patel:2016fam}. 

\section{Nuclear form factor}

To calculate the WIMP-nucleon scattering cross section, one needs following nuclear form factors: $\langle N| m_q \bar q q | N \rangle = m_N^{} f_{T_q}^N$, $(q=u,d,s)$, $ \langle N | -{9 \alpha_s \over 8\pi } G_{\mu\nu}^a G^{a\mu\nu} | N \rangle = m_N f_{T_g}^N $, $\langle N | {\cal O}_{\mu\nu}^q | N\rangle ={1\over m_N} (p_\mu^N p_\nu^N -{1\over 4} m_N^2 g_{\mu\nu}^{} ) (q^N(2) + \bar q^N(2))$,  where $m_N$ is nucleon mass, $f_{T_q}^N$ and $f_{T_g}^N$ are form factors taken from micrOmegas~\cite{Belanger:2018mqt}, $q^N(2) $ and $\bar q^N (2)$ are the second momentum for quark distribution functions evaluated at  $\mu=m_Z$ by using CTEQ PDF~\cite{Pumplin:2002vw}. Specific inputs are~\cite{Belanger:2018mqt}
\begin{eqnarray}
f_{T_u}^p=0.0153\; , \hspace{1cm} f_{T_d}^p=0.0191\; , \hspace{1cm} f_{T_s}^p=0.0447\; ,  \\
f_{T_u}^n=0.0110\; , \hspace{1cm} f_{T_d}^n=0.0273\; , \hspace{1cm} f_{T_s}^n=0.0447\; ,
\end{eqnarray}
$f_{T_g}^N =1-\sum_{q=u,d,s} f_{T_q}^N$, and~\cite{Abe:2018emu,Pumplin:2002vw}
\begin{eqnarray}
u^p(2)=0.220\; , \hspace{0.3cm}d^p(2)=0.110\; , \hspace{0.3cm}s^p(2)=0.026\; , \hspace{0.3cm}c^p(2)=0.019\; , \hspace{0.3cm}b^p(2)=0.012\; ,  \\
\bar u^p(2)=0.034\; , \hspace{0.3cm}\bar d^p(2)=0.036\; , \hspace{0.3cm}\bar s^p(2)=0.026\; , \hspace{0.3cm}\bar c^p(2)=0.019\; , \hspace{0.3cm}\bar b^p(2)=0.012\; . 
\end{eqnarray}

\end{document}